# Concept of a fast neutron detector based on $^{10}$B-RPCs

A. Morozov[1,*], L.M.S. Margato[1], A. Blanco[1], D. Galaviz[2,3]

[1]LIP-Coimbra, Departamento de Física, Universidade de Coimbra, Rua Larga, 3004-516 Coimbra, Portugal

[2]LIP-Lisboa, Av. Prof. Gama Pinto 2, 1649-019, Lisbon, Portugal

[3]Faculty of Sciences, University of Lisbon, Campo Grande, 1749-003, Lisbon, Portugal

[*]Email: andrei@coimbra.lip.pt

## Abstract

We propose an alternative approach for the detection of fast neutrons in the energy range from $10^{-4}$ to 5 MeV based on $^{10}$B-RPCs (hybrid double-gap Resistive Plate Chambers with $^{10}$B$_4$C neutron converters) surrounded by a polyethylene moderator. The detection efficiency as a function of the neutron energy is obtained in Monte Carlo simulations performed with the Geant4 toolkit for several detector configurations. The thickness of the neutron converter is optimized for the maximum efficiency. The results show that for this type of detector it is possible to obtain an average detection efficiency larger than 50% with a weak dependence on the neutron energy. The dark count rate and gamma sensitivity are also discussed.

## 1. Introduction

Fast neutron detectors are widely used in nuclear physics experiments including studies of elastic and inelastic neutron scattering, charge-exchange reactions, photonuclear reactions, neutron-induced fission, and reactions of radioactive nuclei [1]. Some of these applications do not rely on the measurements of the properties of the emitted particle, but rather on the counting of the number of the emitted neutrons. This is the case for the studies of alpha particle induced reactions in low-background dark matter experiments [2], fission products from nuclear reactors [3] and β-delayed neutron emitters [4]. The latter has important implications in operation of power nuclear reactors as well as in nucleosynthesis in nuclear astrophysics [5].

The state-of-the-art fast neutron detector concept for nuclear physics studies is based on the $^3$He-filled proportional counters ($^3$He tubes) embedded in a moderator. Several detection systems were constructed over the past years: see, e.g., [6-8]. Currently the most sensitive fast neutron detector has been build in the frame of the BRIKEN collaboration [9] with more than 70% average detection efficiency in the energy range from $10^{-4}$ to 5 MeV. A comparison of the detection efficiencies of the fast neutron detectors developed for several major experiments can be found in [10]. Since the energy spectrum of the emitted neutrons is often known with a limited precision, the detection efficiency should be as flat (independent of the neutron energy) as possible to minimize the uncertainties in the measured quantities, giving another specification parameter of the detector, the response flatness.

The BRIKEN detector operates with a very large number of $^3$He tubes (up to 179 [9]). The global supply crisis of $^3$He [11] makes purchase of large quantities of that isotope difficult and costly.

Therefore, considering the next generation of fast neutron detectors, it is preferable to find a suitable detection technology not relying on $^3$He.

Recently we have introduced [12] and experimentally validated [13, 14] the concept of a $^{10}$B-RPC thermal neutron detector. The sensitivity of the detector to thermal neutrons is provided by $^{10}$B$_4$C neutron converters (see e.g. [15-17]), deposited on the aluminium cathodes of the hybrid double-gap Resistive Plate Chambers (RPC, see e.g. [18]). The thickness of both electrodes and the gas gap are ≤0.5 mm each, allowing to build very thin detectors. The surface area of an RPC can be quite large (~1 m$^2$), as it was already demonstrated in numerous applications of RPCs for triggering and tracking in High Energy Physics, for example in the ATLAS [19] and CMS [20] experiments at the Large Hadron Collider. $^{10}$B-RPCs operate with ~2 kV potential difference between the cathode and the outer surface of the resistive anode. Despite a strong potential difference in a narrow gas gap, the resistive material of the anode gives the detector an intrinsic discharge protection mechanism.

The detection efficiency of a single $^{10}$B-RPC to normal incidence thermal neutrons is only about 5% [12], however, detectors composed of a stack of $^{10}$B-RPCs can provide >50% detection efficiency, as demonstrated in simulations [12] and experimentally [14]. Manufacture of large area $^{10}$B$_4$C coatings on aluminium substrate is already a well-established technology. For example, the Detector Coatings Workshop of the European Spallation Source is mass-producing such coatings (on the order of hundreds of square meters) for the neutron instruments to be operated at that facility [15-17].

In this study we explore a concept of a $^{10}$B-RPC based fast neutron detector. Small thickness of $^{10}$B-RPCs and the possibility to make them of essentially arbitrary area allow to effectively interlace $^{10}$B-RPCs with thin moderator layers. We propose several detector configurations, inspired by the state-of-the-art fast neutron detectors based on the $^3$He technology, and characterize their performance using Monte Carlo simulations. This study is focused on the analysis of the detection efficiency as a function of the neutron energy, but we also discuss the intrinsic background level and gamma sensitivity of such detectors.

## 2. Methods

### 2.1. Hybrid double-gap $^{10}$B-RPCs

The concept of a hybrid double-gap $^{10}$B-RPC (see figure 1) has been introduced and optimized in our previous simulation and experimental studies [12-14, 21]. Here we assume the following design: the RPC cathode is a 0.35 mm thick aluminium plate, which is positioned between two resistive anodes (soda lime glass, 0.4 mm thick), leaving a 0.35 mm wide gas gap between the cathode and each anode. The uniform width of the gas gaps is ensured by placing spacers (typically nylon monofilaments) with the pitch of a few centimeters. The back surface of the anodes (in respect to the cathode) is lined with a 0.1 mm thick layer of a conductive ink to uniformly distribute the electric potential. The working gas is $C_2H_2F_4$ at atmospheric pressure. A potential difference of about 2 kV is applied between the cathode and the anodes.

The cathodes are lined on both sides with a thin layer of $^{10}$B$_4$C neutron converter. The optimal layer thickness is on the order of 1 μm [12], resulting from a trade-of: the thicker is the converter layer, the higher is the probability for neutron capture by $^{10}$B, but also the lower is the probability for the products of the $^{10}$B(n, α)$^{7}$Li nuclear reaction to reach the gas gap and generate primary ionization. In the presence of a strong electric field, the primary ionization is followed by Townsend avalanches resulting in the induction of detectable signals in the RPC cathodes.

The detection efficiency of one double-gap $^{10}$B-RPC to thermal neutrons is quite low: for normal incidence it is about 5 to 10% depending on the neutron energy [12]. The detection efficiency increases with the decrease of the angle between the incidence direction and the RPC surface: for thin layers it scales as inverse sine of that angle. Therefore, to make a detector with >50% detection efficiency, it is either composed of a stack of $^{10}$B-RPCs with tens of converter layers at normal incidence, or the neutron incidence angle is chosen to be close to ~5 degrees [12, 14].

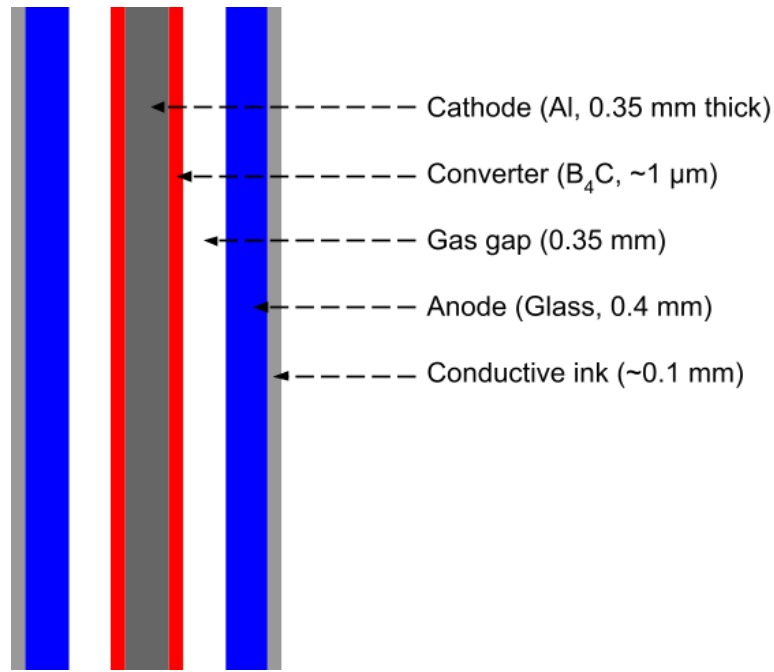

Figure 1. Schematic drawing of a double gap $^{10}$B-RPC: side view, not to scale. The total thickness of a single unit is about 2 mm.

## 2.2. Fast neutron detector based on $^{10}$B-RPCs

As $^{10}$B-RPCs have very low sensitivity to fast neutrons (see section 4.3), a neutron moderator has to be added to the design. We have considered several possible approaches, which are summarized in this chapter. For all of them we have considered HDPE (high density polyethylene) as the moderator material.

The first approach benefits from the planar geometry of the $^{10}$B-RPCs: sheets (or slabs) of the moderator material are introduced between neighboring $^{10}$B-RPCs. As an example inspired by the geometry of the BRIKEN detector [9], a fast neutron $^{10}$B-RPC detector can be designed as shown in figure 2. The moderator is a cube of 800 mm side size. Two rectangular (100 mm side size) vacuum channels cross the detector in the middle and it is assumed that an isotropic neutron source is positioned in the geometric center. $^{10}$B-RPCs of two sizes are introduced in the design: the large-size ones (800 × 800 mm$^2$) are installed as equally-spaced layers above and below the vacuum channels, while the smaller ones (100 × 350 mm$^2$) are positioned perpendicularly to the large $^{10}$B-RPCs in the space around the channels. A large pitch between the $^{10}$B-RPC planes in figure 2 is chosen for convenience of presentation. An optimal value of the pitch has to be determined as a compromise between the requirements on the detection efficiency and the maximum allowed $^{10}$B-RPC area (see section 3).

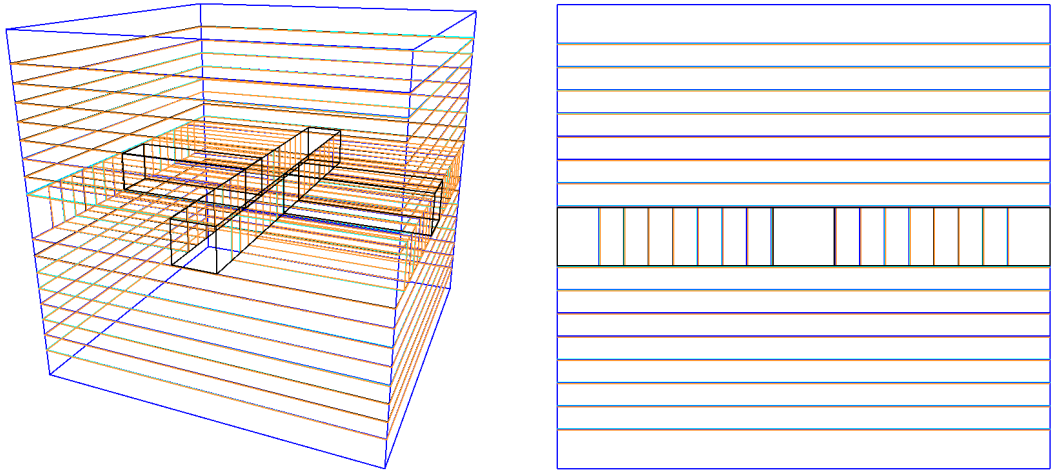

Figure 2. Perspective (left) and the side view (right) of a conceptual design of a fast neutron detector based on $^{10}$B-RPCs with a cubic moderator and two vacuum channels crossing the detector in the center.

Another approach, inspired by the geometry of the BELEN detector [22], is to have a cylindrical moderator with a vacuum channel along the axis and to install $^{10}$B-RPCs at ~10 degrees with respect to the radial direction (see figure 3). Such an angle in the $^{10}$B-RPC positioning is needed to prevent neutrons from escaping along the gas gap channels as an isotropic neutron source is again assumed to be in the geometric center.

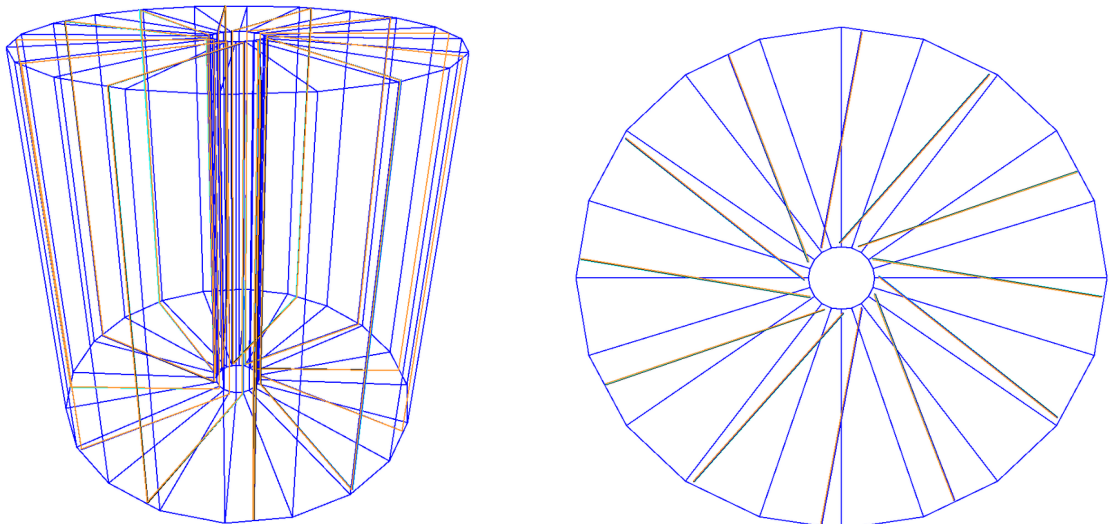

Figure 3. Perspective (left) and the top view (right) of a conceptual design of a fast neutron detector based on $^{10}$B-RPCs with a cylindrical moderator and a vacuum channel along the moderator's axis. The radial blue lines are a part of the graphical representation of the cylindrical moderator.

In this approach the separation between the neighboring $^{10}$B-RPCs increases with the distance from the cylinder's axis, which should lead to a decrease in the detection efficiency with the neutron energy. This issue is explored in section 3.4 and a modification of the detector geometry is proposed based on the simulation results.

Finally, the third approach discussed in this paper is to combine several $^{10}$B-RPCs in narrow "assemblies", which then can be installed inside the moderator in a way similar to the one used with $^{3}$He tubes. Each assembly is encased in a thin-walled aluminium box and the neighboring $^{10}$B-RPCs are separated with a few millimeter thick moderator layers (see figure 4, left). The length of an assembly is quite flexible and should be defined based of the size and shape of the moderator. Figure 4 (right) shows an example of the positioning pattern of the assemblies inside the moderator and further details are given in sections 3.3 and 3.5.

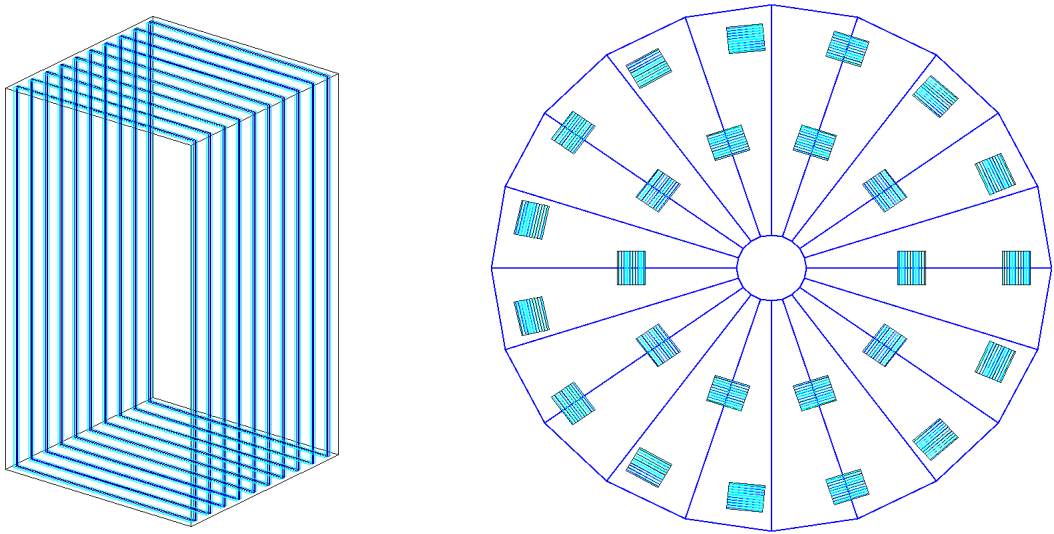

Figure 4. Left: A perspective view of a $^{10}$B-RPCs assembly, featuring 10 double-gap $^{10}$B-RPCs. The assembly is encapsulated in a thin wall aluminium box, and the $^{10}$B-RPCs are interlaced with thin moderator layers. The length (vertical scale here) can be made significantly larger than the size in the perpendicular directions. Right: An example of the positioning pattern of the assemblies inside a cylindrical moderator (top view). The long side of the assemblies is aligned parallel to the moderator's axis.

## 2.3. Simulation tools

All simulations in this study were performed with the Geant4 toolkit [23-25] version 10.7.2. The detector optimization and comparison of the performance characteristics for different detector designs were conducted using the QGSP_BIC_HP reference physics list. Several cross-tests have also been performed using the Geant4's thermal libraries (G4NeutronHPThermalScattering), giving a few percent lower detection efficiency values as described in section 3.6. The results obtained with those libraries should, theoretically, be more realistic as the code takes into account the molecular bonds of hydrogen in HDPE for the simulation of neutron elastic scattering. As reported in [9], several errors were fixed in the interpolation routines in version 10.0 and the results of Geant4 simulations with the following Geant4 versions were essentially the same as the ones provided by the MCNPX code.

However, we have decided against using the thermal libraries in the bulk of this study due to the following reasons. Recently we have reported a major problem in the generation of the neutron scattering direction (report #2290 in the Geant4 problem tracking system [26]), confirming a problem reported more than five years ago. This error has been fixed in the recent version 10.7.1. Since according to the Geant4 developers there is currently no one in charge of upkeeping and developing this part of the code, we concluded that it is a less risky option to use the mainstream neutron library.

ANTS2 toolkit version v2.46 [27] served as the front-end for simulations and was used to configure the detector's geometry, set-up the neutron source, organize scoring in Geant4, log energy deposition and analyze the results.

## 2.4. Simulation procedures

The simulations were performed with an isotropic point source of monochromatic neutrons situated in the geometric center of the detector. Typically, $4\times10^4$ neutrons were simulated for each energy. To simplify comparison with already published results for $^3$He-based detectors [9, 28], we have used the following discrete neutron energies:

$E_n$ = {$10^{-4}$, $10^{-3}$, $10^{-2}$, $10^{-1}$, 1, 2, 3, 4, 5} in MeV.

The production cut of Geant4 was set to 0.01 mm. The maximum tracking step was configured to 0.01 mm in the gas gaps and to $10^{-5}$ mm inside the neutron converters using G4StepLimiterPhysics.

A neutron event was considered detected if the neutron capture reaction products deposited at least 0.1 MeV in a gas gap. This energy threshold gives a good match in the simulated and experimental neutron detection efficiencies for a multi-layer $^{10}$B-RPC thermal neutron detector characterized at neutron beam [21].

## 2.5. Performance parameters of fast neutron detectors

According to the literature on the development of fast neutron detectors for the measurements of the probability of β-delayed neutron emission (see, e.g., [9, 22]), the detectors need to be sensitive to neutrons in the energy range from 100 keV to ~5 MeV, have the detection efficiency (DE) of 50% or higher, and the DE should be as flat as possible in respect to the neutron energy. The performance of a fast neutron detector is typically characterized by the following two parameters:

(1) *Average detection efficiency*, given by the mean value of the detection efficiencies obtained with monochromatic neutrons for a set of discrete energies, and

(2) *Flatness*, given by the ratio of the maximum and minimum values of the detection efficiencies obtained with monochromatic neutrons for the same set of discrete energies.

These two parameters are used in sections 3.1 – 3.5 to compare the performance of the different designs of the fast neutron detector based on $^{10}$B-RPC. Two additional parameters, the intrinsic background counting rate and the detector's gamma sensitivity, both of which can be critical in evaluation of the suitability of the detection technology for a given type of measurements, are discussed in sections 4.2 and 4.3.

# 3. Results

## 3.1. Cubic moderator with regular $^{10}$B-RPC spacing

The first configuration of the fast neutron detector, introduced in section 2.2 (figure 2) has a cubic moderator with $^{10}$B-RPCs positioned with regular spacing. Simulated detection efficiency as a function of the neutron energy is shown in figure 5 for several values of the pitch in $^{10}$B-RPC positions inside the moderator.

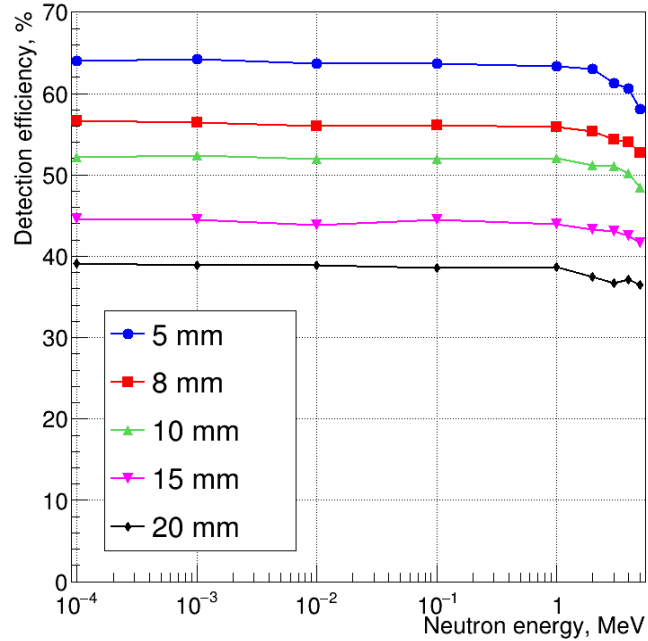

Figure 5. Detection efficiency as a function of the neutron energy for the configuration with the cubic moderator and regular pitch in $^{10}$B-RPCs positions inside the moderator. The data are shown for several pitch values from 5 to 20 mm. The converter layer thickness is 0.7 μm. The statistical uncertainties are smaller than the size of the markers.

Table 1 lists the total area of the RPC cathode, the average detection efficiency (DE) and the flatness (see section 2.5) for the same values of the pitch which appear in figure 5.

| Pitch, mm | Overall cathode area, m$^2$ | Average DE, % | Flatness |
|---|---|---|---|
| 5 | 99.4 | 62.4 | 1.11 |
| 8 | 61.1 | 55.3 | 1.08 |
| 10 | 49.7 | 51.2 | 1.08 |
| 15 | 32.7 | 43.5 | 1.07 |
| 20 | 24.1 | 38.0 | 1.07 |

Table 1. Parameters of the detector with the cubic moderator and regular spacing of $^{10}$B-RPCs for several values of the pitch and the B$_4$C thickness of 0.7 μm. To get the total area of the neutron converter, multiply the cathode area by two. The statistical uncertainty in the DE and the flatness data are less than 0.2 and 0.01, respectively.

The data presented in figure 5 and table 1 were obtained with the converter thickness of 0.7 μm. The dependence of the average detection efficiency on the converter thickness was investigated for the pitch values of 8 and 20 mm. The results are shown in figure 6, suggesting that the highest efficiency is reached with the layer thickness in the range of 0.7 – 1.0 µm and that the optimal thickness slightly increases with the pitch. The curves show broad peaks, which is good from the practical point of view as it allows to set quite large tolerances for the $B_4C$ deposition process. The simulations have also shown that the flatness does not change with the converter thickness.

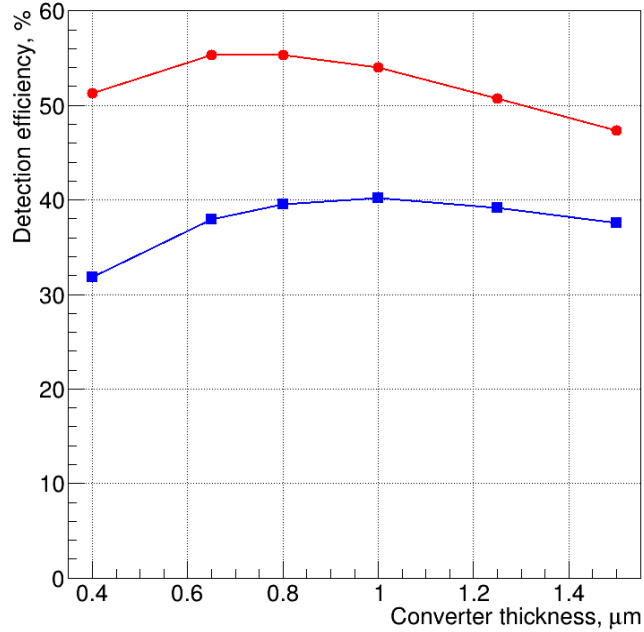

Figure 6. Average detection efficiency as a function of the converter thickness for the configuration with the cubic moderator and regular spacing of $^{10}$B-RPCs. The pitch values are 8 mm (red round dots) and 20 mm (blue square dots).

The relative fractions of all non-negligible end-of-track scenarios for the primary neutrons are listed in tables 2 (5 mm pitch) and 3 (20 mm pitch). The *total inelastic* column represents the fraction of neutrons captured by $^{10}$B. Since not all neutrons captured by $^{10}$B result in energy deposition in the gas gap above the detection threshold, the *detected* fraction is smaller. The *capture* column represents the fraction of neutrons captured in the moderator. The *total escape* column represents the fraction of neutrons escaping the detector, and the *full energy escape* column gives the fraction of neutrons which exit the moderator with the energy loss smaller than 0.001 of the initial value. Note that the total solid angle fraction of the vacuum "ports" is 2.0% for the neutron source at the geometry center. The data in these two tables are obtained with the converter thickness of 0.9 µm.

| Neutron energy, MeV | Total inelastic,% | Detected, % | Capture, % | Total escape, % | Full energy escape, % |
|---|---|---|---|---|---|
| $10^{-4}$ | 81.9 | 61.6 | 12.3 | 5.8 | 2.1 |
| $10^{-2}$ | 81.3 | 61.4 | 12.3 | 6.5 | 2.1 |
| 1 | 81.3 | 61.2 | 12.1 | 6.7 | 2.4 |
| 3 | 79.4 | 58.8 | 11.1 | 9.5 | 3.1 |
| 5 | 77.1 | 56.0 | 9.6 | 13.3 | 3.6 |

Table 2. End-of-track scenario fractions for 5 mm pitch.

| Neutron energy, MeV | Total inelastic,% | Detected, % | Capture, % | Total escape, % | Full energy escape, % |
|---|---|---|---|---|---|
| $10^{-4}$ | 54.8 | 41.5 | 38.1 | 7.1 | 2.1 |
| $10^{-2}$ | 53.8 | 40.7 | 38.7 | 7.4 | 2.0 |
| 1 | 53.4 | 40.2 | 39.8 | 6.8 | 2.1 |
| 3 | 52.3 | 38.7 | 39.5 | 8.2 | 2.8 |
| 5 | 52.2 | 37.9 | 37.3 | 10.5 | 3.2 |

Table 3. End-of-track scenario fractions for 20 mm pitch.

We have also analyzed the dependence of the relative "load factors" of $^{10}$B-RPCs on the distance from the detector center for different neutron energies. The $^{10}$B-RPCs were numbered from 1 (closest to the detector center) and up to 70, which is the maximum index for the configuration with the 5 mm pitch. Figure 7 shows the contributions of the $^{10}$B-RPCs with the same index to the total detection efficiency as a function of the index for several neutron energies. Note that for each energy the integral under the corresponding curve gives the value shown in table 2 (*Detected* column).

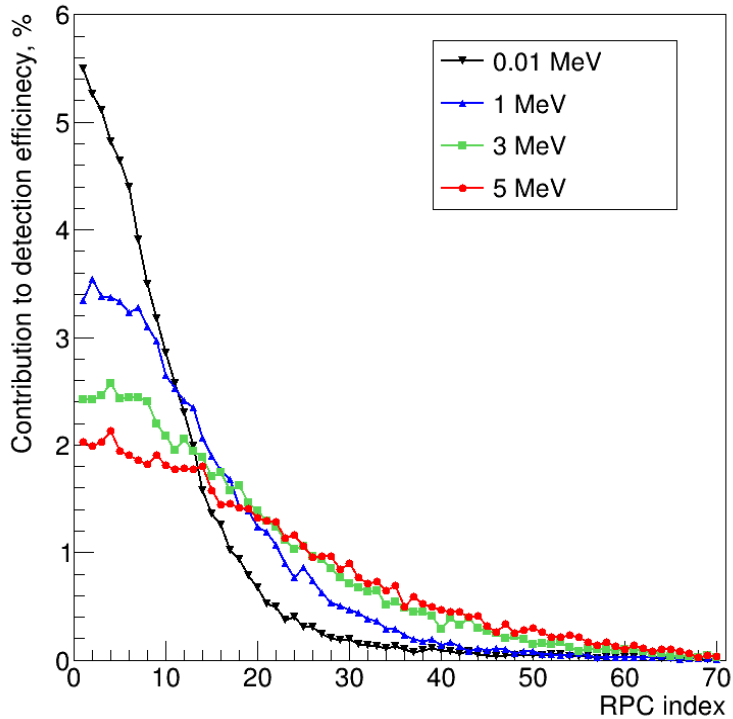

Figure 7. Contributions to the detection efficiency as a function of the $^{10}$B-RPC index for the configuration with the cubic moderator and 5 mm pitch for several neutron energies.

## 3.2. Cubic moderator with a variable $^{10}$B-RPC pitch

Figure 7 shows that the contribution of individual $^{10}$B-RPCs strongly decreases with increase of the index, even for the neutron energies above 1 MeV. A possible modification of the design, aimed at the reduction of the total area of $^{10}$B-RPCs, can be in having more dense packing of $^{10}$B-RPCs close to the center compared to the periphery. For example, the first 10 $^{10}$B-RPCs from the detector center can be installed with a 5 mm pitch, the next 17 with a 8 mm one and the last 9 with a 20 mm pitch (see figure 8, left). This approach results in an area of 51.2 m$^2$, which is approximately half the value for the configuration with the regular pitch of 5 mm.

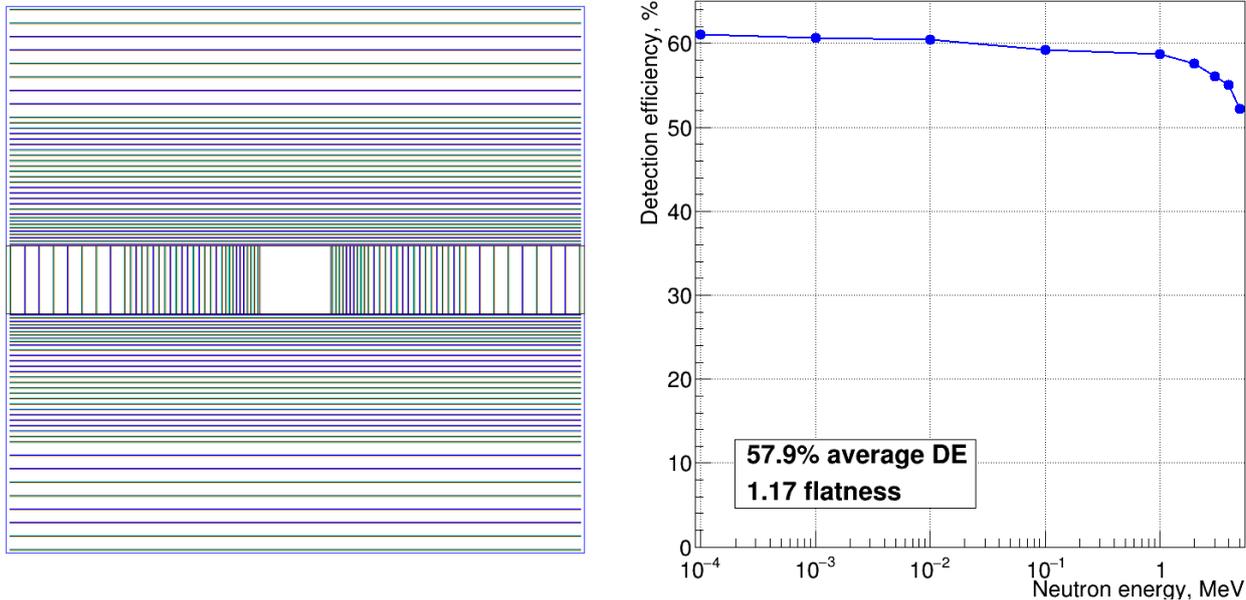

Figure 8. Side view (left) and the detection efficiency as a function of the neutron energy for the configuration with the cubic moderator and $^{10}$B-RPCs placement with a variable pitch.

The simulated detection efficiency as a function of the neutron energy for this configuration with 0.9 μm converter thickness is shown in figure 8 (right). The average detection efficiency is 57.9%, which is only slightly worse than the value of 62.4% obtained for the configuration with the regular pitch of 5 mm. However, quite predictably, the flatness of 1.17 is significantly worse compared to the value of 1.11 for the configuration with the regular pitch.

The end-of-track scenario fractions for 5 MeV neutrons are 71.7% total inelastic, 52.6% detected, 17.5% capture in the moderator, 10.8% total escape and 3.2% full escape. As expected, compared with the corresponding data for the configuration with the regular pitch of 5 mm (table 2), there is a noticeable redistribution of the scenario fractions from the total inelastic to the capture in the moderator.

## 3.3. Cubic moderator with $^{10}$B-RPC assemblies

As discussed in section 2.2, it is also possible to use narrow box-shaped $^{10}$B-RPC assemblies positioned inside the moderator in a way similar to the one used with $^3$He tubes. Here we have characterized the detector performance with such assemblies each composed of ten 50 mm wide and 348 mm long $^{10}$B-RPCs, encapsulated in thin wall aluminium boxes with the dimensions of 52 × 40 × 350 mm$^3$. The boxes are "filled" with HDPE and $^{10}$B-RPCs are positioned inside the boxes with a 4 mm pitch. The cross-section area of an assembly is similar to the one of a two-inch $^3$He tube, and the RPC width is large enough from the practical point of view, considering deposition of the converter layer and manufacturing of the assembly.

The assemblies are positioned inside a 820 mm side cube of HDPE as depicted in figure 9. The distance between the neighboring assemblies in the horizontal and vertical directions (see figure 9, left) is 15 mm, with the exception of the two middle rows, which are separated in the vertical direction by 10.5 mm. There are 236 assemblies and the total cathode area is 41.2 m$^2$. Note that the number of assemblies can be reduced nearly by half if assemblies of two lengths (348 mm long ones for the "cross" area and ~700 mm long ones for the rest of the detector) are used.

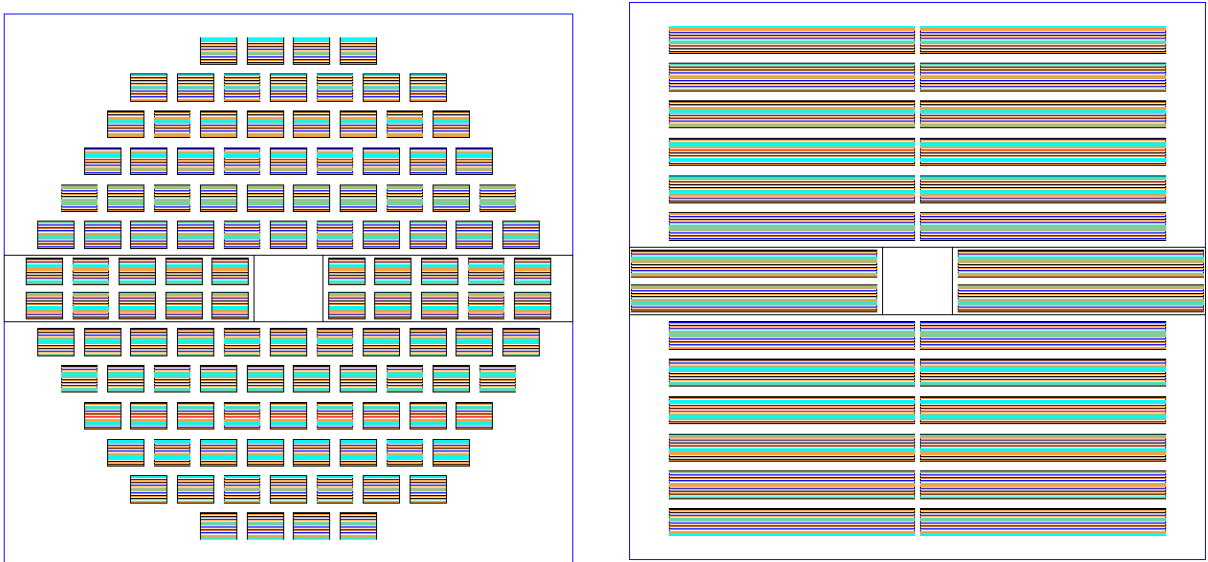

Figure 9. Front and side view of the detector configuration with a cubic moderator and $^{10}$B-RPC assemblies of 52 × 40 × 350 mm$^3$.

The detection efficiency as a function of the neutron energy is shown for the converter thickness of 0.7 μm in figure 10 (left). The average detection efficiency is 52.3% and the flatness is 1.08. The average detection efficiency as a function of the converter thickness is given in figure 10 (right). Note that at 5 MeV the detection efficiency is still quite high: 49.8%. The end-of-track scenario fractions for that energy are 64.7% total inelastic, 23.8% capture in the moderator, 11.6% total escape and 3.4% full escape. There is even larger redistribution of the scenario fractions from the total inelastic to the capture in the moderator compared to the configuration with the variable pitch (section 3.2), while the escape fractions are very similar for all the considered configurations.

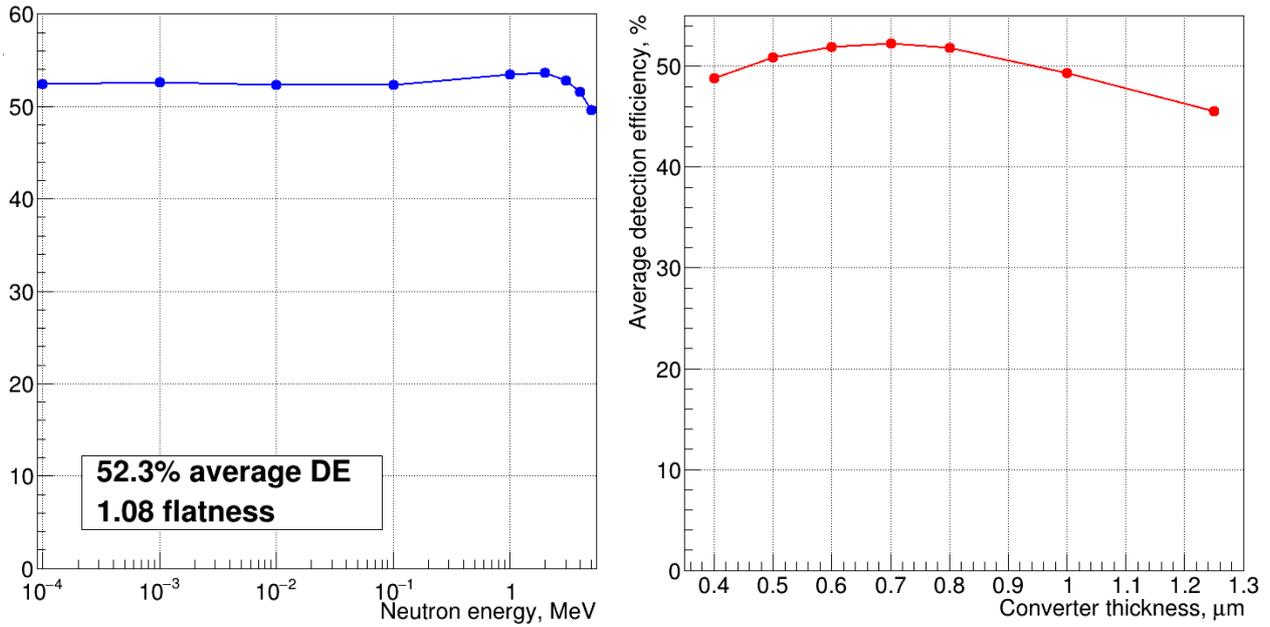

Figure 10. Detection efficiency as a function of the neutron energy (left) and the average detection efficiency as a function of the converter thickness (right) for the detector configuration with the cubic moderator and $^{10}$B-RPCs assemblies.

The performance of the detector in this configuration is quite good: >50% average detection efficiency combined with high flatness, while the total cathode area is only half of the value for the configuration with the regular pitch of 5 mm. This configuration is also more attractive from the practical point of view due to the possibility to insert/replace individual assemblies not affecting the rest of the detector. Note that further optimization involving positioning of the assemblies in the moderator, similar to the one described in [28] can be performed for a concrete experimental requirements and a given limit on the total $^{10}$B-RPC area. Obviously, the larger is the acceptable area, the better values of the average DE and flatness can be obtained.

### 3.4. Cylindrical moderator with $^{10}$B-RPC planes

Another configuration of the detector, introduced in section 2.2 has a cylindrical moderator. Here we have considered a HDPE cylinder with 800 mm outer and 100 mm inner diameters. The $^{10}$B-RPCs planes are inserted parallel to the cylinder axis with a 10 degree angle to the radial direction (see figure 11, left). The number of $^{10}$B-RPC planes is 64, which was selected to have the minimum distance between the neighboring planes of about 5 mm. The total cathode area for this configuration is 18.2 m$^2$.

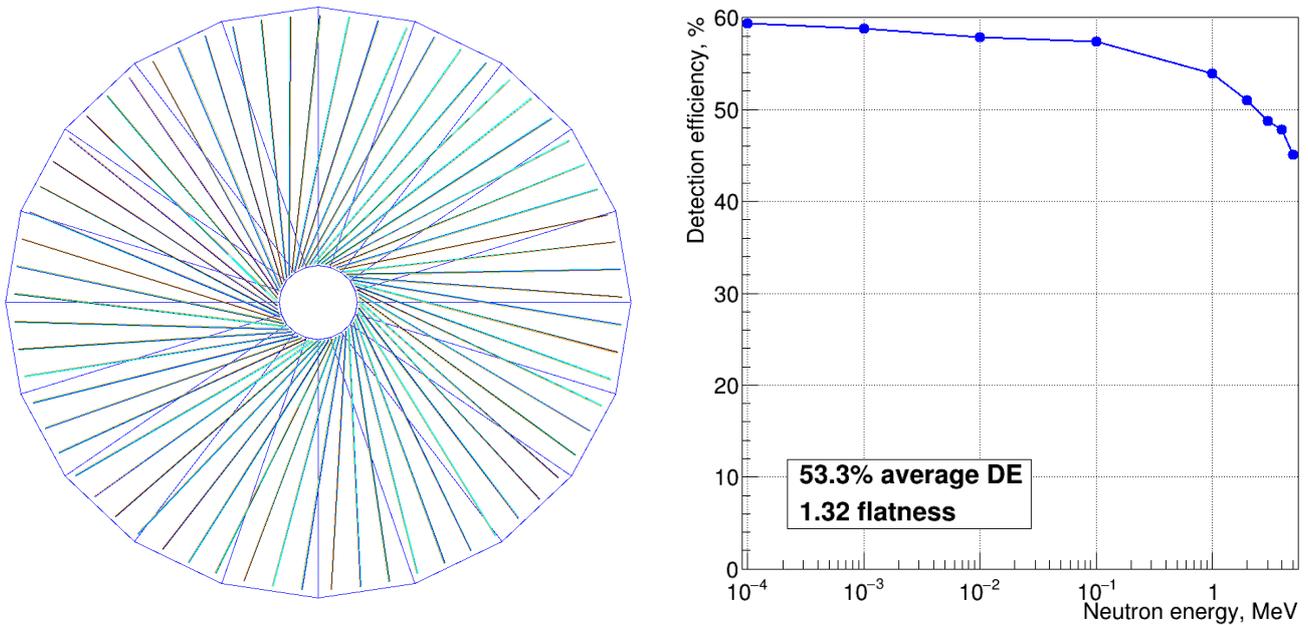

Figure 11. Top view of the geometry (left) and the detection efficiency as a function of the neutron energy (right) for the configuration with the cylindrical moderator and one ring of 64 $^{10}$B-RPCs planes.

The detection efficiency as a function of the neutron energy is shown in figure 11 (right) for $^{10}$B-RPCs with 0.7 μm thick converters. The average efficiency is 53.3%, which is a good result for such a small cathode area. However, the detection efficiency strongly reduces with the neutron energy, resulting in a bad flatness value of 1.32. This is an expected result as the separation between the neighboring $^{10}$B-RPCs increases with the radial distance and reaches 39 mm at the periphery of the moderator.

One way to counter such a large increase in the separation between the $^{10}$B-RPCs is to use narrower $^{10}$B-RPCs and position them in several rings, as it is shown in figure 12 (left). In this configuration there are three rings with 65, 100 and 150 $^{10}$B-RPCs. The detection efficiency indeed improves at high neutron energy (figure 12, right) and the flatness reaches a value of 1.23. However, the price of this improvement is nearly double increase in the cathode area (from 18.2 to 27.7 m$^2$).

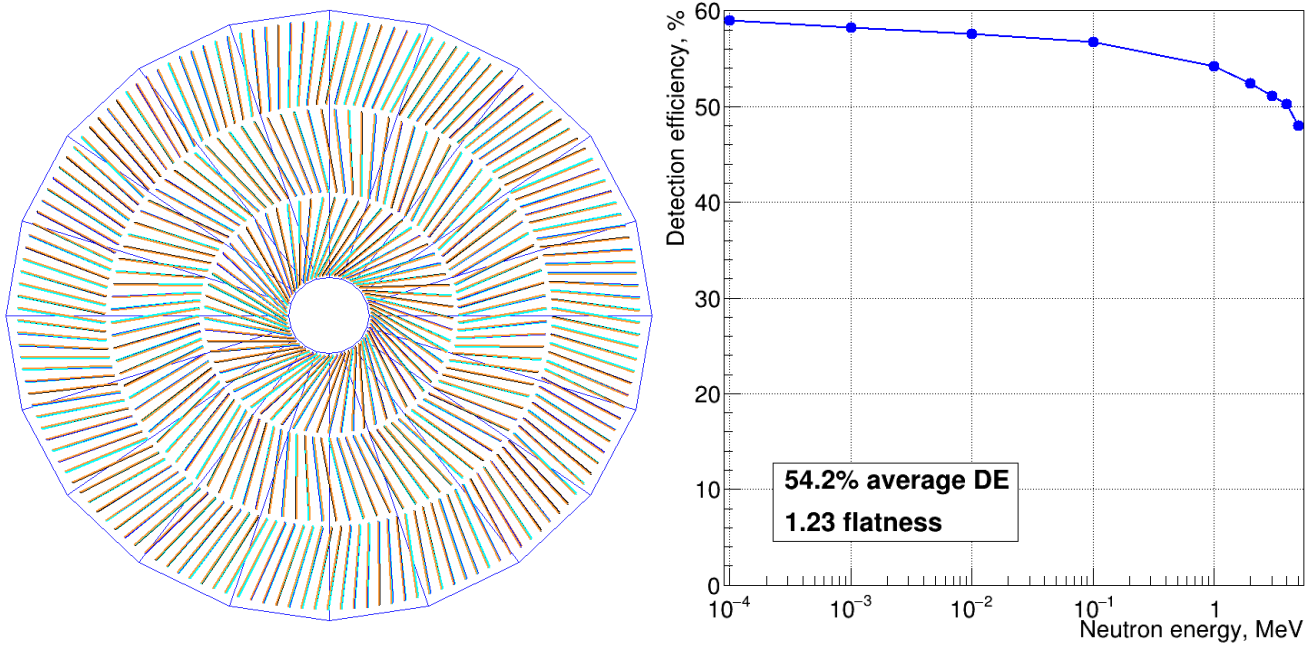

Figure 12. Top view of the geometry (left) and the detection efficiency as a function of the neutron energy (right) for the configuration with the cylindrical moderator and three rings of $^{10}$B-RPCs (65, 100 and 150 planes for the inner, middle and outer rings, respectively).

The end-of-track scenario fractions for 5 MeV neutrons are 63.8% total inelastic, 48.2% detected, 29.5% capture in the moderator, 6.8% total escape and 1.6% full energy escape. The smaller escape fractions compared to the configuration with the cubic moderator is due to the geometry: there is only one vacuum channel in this configuration, and its area is smaller.

Note that if good flatness is mandatory while the requirement on the average detection efficiency is not strict, the number of $^{10}$B-RPCs in the inner ring can be reduced. As the result, the detection efficiency decreases more for the lower energies, and flatness improves. For example, with a quite drastic reduction of the number of the inner ring $^{10}$B-RPCs from 65 to 33, the flatness improves to 1.08, however, the average detection efficiency reduces to 42.9%.

## 3.5. Cylindrical moderator with $^{10}$B-RPCs assemblies

We have also considered a configuration with $^{10}$B-RPC assemblies positioned in a cylindrical moderator. Increasing the length of the assemblies defined in section 3.3 from 350 mm to 700 mm and placing them in the pattern shown in figure 13 (left), we obtain a configuration with 108 assemblies and the total cathode area of 43.1 m$^2$. Note that the inner ring assemblies are rotated 90 degrees in order to maximize the packing density.

The detection efficiency as a function of the neutron energy is shown for this configuration in figure 13 (right). The average detection efficiency and flatness were found to be 54.4% and 1.06, respectively. These values are slightly better than the ones obtained in section 3.3. An analysis of the dependence of the average detection efficiency on the converter layer thickness gives essentially the same curve as the one shown in figure 10 (right): the maximum efficiency is reached at 0.7 μm.

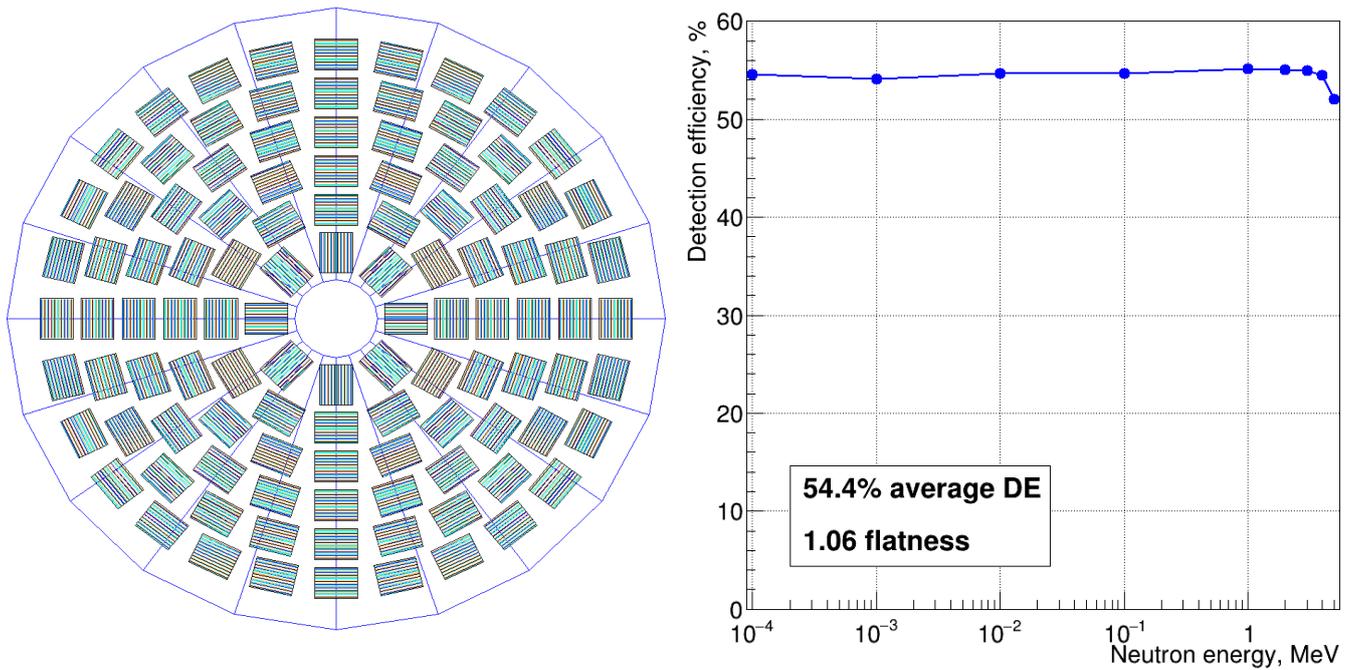

Figure 13. Top view of the geometry (left) and the detection efficiency as a function of the neutron energy (right) for the configuration with the cylindrical moderator and 108 $^{10}$B-RPC assemblies.

The end-of-track scenario fractions for 5 MeV neutrons are 66.6% total inelastic, 52.6% detected, 26.9% capture in the moderator, 6.6% total escape and 1.3% full energy escape.

We have also analyzed the contribution of the outer rings of $^{10}$B-RPC assemblies to the total detection efficiency. For this purpose, in addition to the configuration with 6 rings of the assemblies (figure 13), designated below as *Six rings*, two new configurations were considered. In the first one, designated as *Five rings*, the most outer ring of the assemblies was removed from the *Six rings* configuration keeping the moderator diameter unchanged. In the second one, an additional ring of 32 assemblies was added to the *Six rings configuration, and the moderator diameter was increased* to 900 mm to accommodate the additional ring (*Seven rings* configuration). The detection efficiency as a function of the neutron energy for these three configurations are shown in figure 14.

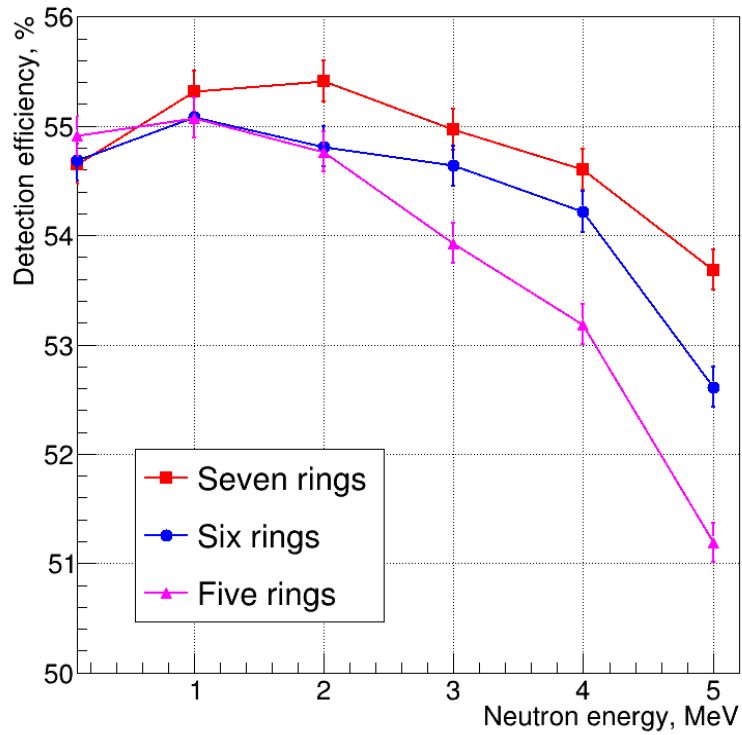

Figure 14. Detection efficiency as a function of the neutron energy for three configurations of the detector with the cylindrical moderator and $^{10}$B-RPC assemblies. The data are shown in the energy range from 0.1 to 5 MeV. For lower energies ($10^{-4}$ – $10^{-2}$ MeV, not shown) the detection efficiencies are the same for these configurations considering the statistical uncertainties.

As expected, for energies ≤ 1 MeV there is no systematic difference, while for 5 MeV neutrons an increase in the number of rings of assemblies gives an improvement in the efficiency of about 1% with each ring. The average DE and flatness values are listed in table 4 together with the total $^{10}$B-RPC cathode area.

| Configuration | Average DE, % | Flatness | Cathode area, m$^2$ |
|---|---|---|---|
| Five rings | 54.1 | 1.07 | 31.9 |
| Six rings | 54.4 | 1.06 | 43.1 |
| Seven rings | 54.7 | 1.03 | 55.9 |

Table 4. Detector characteristics for the configurations with the different number of rings of $^{10}$B-RPC assemblies.

## 3.6 Timing properties

Another characteristic of the detector which is important in calculation of the signal duration is the distribution of the time passed from emission of a neutron to its detection. For the configuration presented in the previous section (108 $^{10}$B-RPC assemblies installed in a 800 mm diameter HDPE moderator), two examples of such distribution are shown in figure 15 for neutron energies of 0.1 and 5 MeV. The neutrons are emitted at zero time. Both datasets very closely resemble an exponential distribution with the decay time of about 0.055 ms and 99% of all neutrons are detected withing 0.3 ms.

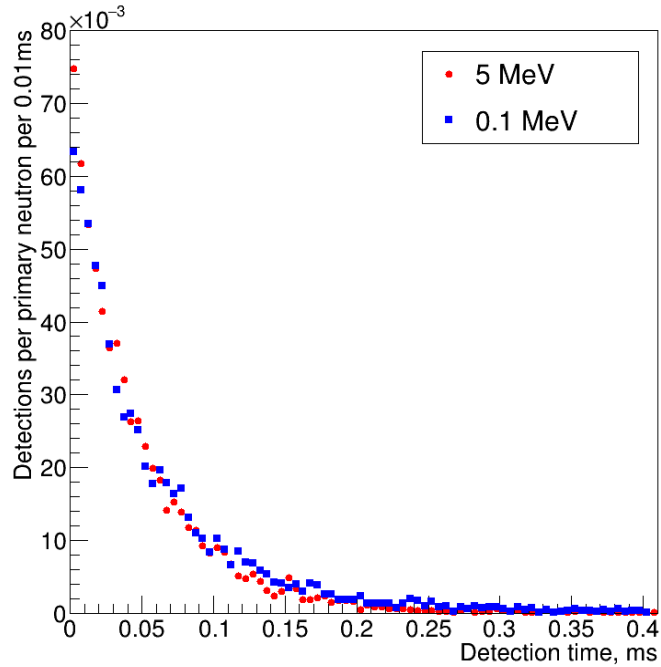

Figure 15. Distribution of the time of detection for neutrons emitted at zero time with 5 MeV (red dots) and 0.1 MeV (blue rectangles) energies.

## 3.7 Results obtained with the thermal libraries of Geant4

As described in section 2.3, the results presented in this study were obtained using Geant4's QGSP_BIC_HP physics list without activation of the thermal libraries. However, for one detector configuration we have made a cross-test between simulations performed without and with those libraries. In the former case, the elastic scattering cross-section of hydrogen in HDPE material uses generic hydrogen data without taking into account the molecular bonds in that polymer. In the later case the cross-section is constructed based on experimental data obtained with a polyethylene target.

The results for both approaches are given in figure 16. Activation of the thermal libraries leads to a small reduction (factor of 0.965) in the average detection efficiency. The flatness is not affected.

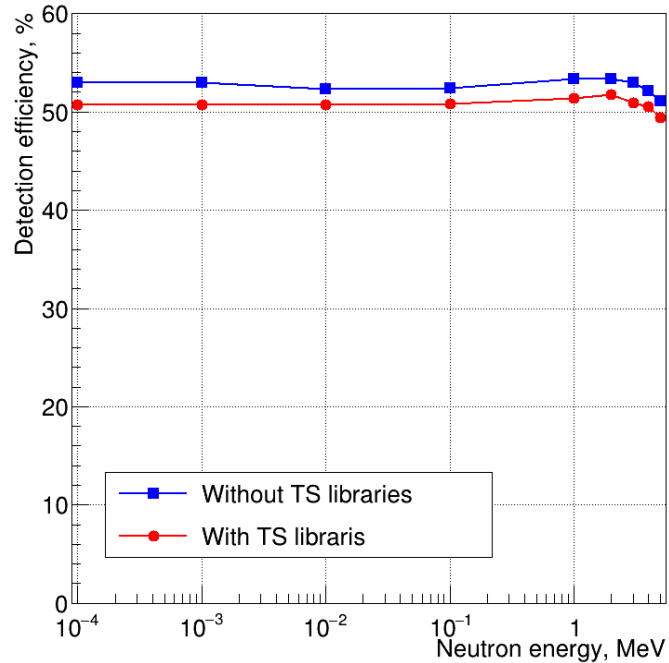

Figure 16. Detection efficiency as a function of the neutron energy for the same detector configuration, simulated with (red round dots) and without (blue rectangles) activation of the Geant4's thermal libraries. The statistical uncertainties are smaller than the size of the markers.

# 4. Discussion

## 4.1. Average detection efficiency and flatness

The results of this study suggest that fast neutron detectors based on the $^{10}$B-RPC technology can reach average detection efficiencies of 50 – 60% with high (~1.05) flatness in the neutron energy range from $10^{-4}$ to 5 MeV. These detection efficiencies are less than the values of 65 – 70% demonstrated by several modifications of the $^3$He-based BRIKEN detector in the same energy range [10] mainly due to the energy loss of the $^{10}$B(n, α)$^7$Li reaction products in the converter layer. The data in table 2 suggests that only about 75% of neutrons captured by $^{10}$B can be detected.

However, $^{10}$B-based technologies have an important practical advantage: a $^3$He-based detector, such as BRIKEN, requires about 150 $^3$He tubes. Due to the ongoing $^3$He supply crisis, it is very difficult to get access to such large number of tubes and even if it is, the cost can exceed one million euros. In contrast, $^{10}$B$_4$C coatings are much more accessible and the cost of deposition of ~1 μm thick $^{10}$B$_4$C layer on aluminium substrate for ~50 m$^2$ is on the order of 50 thousand euros.

Comparing the designs of the $^{10}$B-RPC fast neutron detector analyzed in this study, the ones utilizing the compact $^{10}$B-RPCs assemblies seems to be the most attractive. For the $^{10}$B-RPC cathode area <50 m$^2$ they can provide average detection efficiency >50% and flatness <1.10 in the neutron range up to 5 MeV. In comparison to the designs involving large area $^{10}$B-RPCs, the average detection efficiency is somewhat lower, but there are strong practical advantages based on the modularity of the design, especially considering construction and maintenance of the detectors.

The detector configuration can be further optimized based on the concrete requirements on the average detection efficiency and flatness. However, as demonstrated by figure 14 and table 4, for high-performance configurations, even a small improvement in the flatness can requires quite a significant increase in the $^{10}$B-RPC area.

## 4.2. Dark count rate

Low intrinsic background of a fast neutron detector is an important requirement for its applicability in some areas, such as, for example, the studies of β-delayed neutron emitters [29]. At the current state of the development, the technology based on $^{10}$B-RPCs can be in disadvantage in this respect to the one based on $^{3}$He proportional counters. The first prototypes of $^{10}$B-RPCs characterized in our validation studies show dark count rates on the order of 0.02 cps/cm$^2$ [13]. For a $^{10}$B-RPC with the total cathode area of ~50 m$^2$ the dark count rate extrapolates to about 10 kcps. This value is an order of magnitude larger than the on-beam background of the BRIKEN detector [29, 30]: after application of a neutron-gating technique, it was possible to operate BRIKEN with the background rate of ~0.5 kcps. Also, considering detection of single neutrons with such $^{10}$B-RPC detector and using 0.3 ms measurement time window (see section 3.5), the expected number of dark counts is about 3, which also needs more than an order of magnitude reduction to make single neutron detection feasible.

However, all $^{10}$B-RPCs so far constructed and tested were small area prototypes (<100 cm$^2$) operated on bright neutron beams. The measured dark count rates (a few cps) were negligible in comparison with the neutron count rates and, therefore, no studies aiming to optimize the $^{10}$B-RPC design to reduce this background were yet conducted.

There is no agreement in the scientific community on the exact origin of the RPC dark counts. It is, however, known that unevenness of the electrode surface, the presence of dust particles and contamination of the surface of the metallic cathodes with dielectric layers/films can strongly affect the dark count rate [18, 31]. Therefore, in order to reduce the intrinsic background in the future prototypes, the assembly of $^{10}$B-RPCs should be made in an extra-clean environment and special conditioning routines (operation at an elevated polarization potential with argon instead of the working gas) should be applied to "burn away" the contaminants and "needles" at the surface. Also, benefiting from the fact that $^{10}$B-RPCs are not expensive, it is possible to manufacture an excess of them and perform a selection of the low noise ones.

Another approach to decrease the dark count rate is to reduce the RPC polarization potential. In order to maintain the same detection efficiency, this approach may require to use more sensitive front-end electronics to be able to compensate for the decrease in the gas gain and the resulting reduction in the amplitude of the induced signals.

Finally, radio-pure materials can be considered for manufacturing of the $^{10}$B-RPC components, in particular, aluminium cathodes, as significant amounts of U and Th isotopes may be present in the aluminium ore [32]. These isotopes contribute to the background mainly due to the emission of alpha particles, which are able to reach the RPC gas gap through the converter layer [33].

## 4.3. Gamma sensitivity and "direct" sensitivity to fast neutrons

Simulations of 10$^6$ gammas emitted with the energy of 1 MeV from an isotropic point source situated in the geometric center of the detector described in section 3.5 did not result in any events with the value of energy deposited in a gas gap above the detection threshold. This result suggests that the detector's sensitivity to gamma rays should be quite low, which is also confirmed by the results of our experimental study on the gamma sensitivity of $^{10}$B-RPC thermal neutron detectors: for 1 MeV the sensitivity of a detector prototype was found to be about $10^{-5}$ [34].

A series of simulations was also performed with a detector configuration from section 3.5 in which all $^{10}$B$_4$C converters were removed from the detector geometry. The results show that the detection efficiency in that case stays below 0.01% for all neutron energies. This fact suggests that only a negligible fraction of neutrons deposit an amount of energy above the detection threshold inside the gas gaps directly (due to the nuclear recoil). An important consequence of this fact is that the scenarios in which one neutron deposits energy in several gas gaps can be ignored.

## 5. Conclusions

The results of this study suggest that $^{10}$B-RPC technology has a good potential for fast neutron detectors which require high (and flat) detection efficiency in the energy range up to 5 MeV. Based on $^{10}$B$_4$C neutron converters, this technology should allow to construct high performance detectors for an order of magnitude lower price compared to the current $^3$He-based detectors.

Several configurations analyzed in this study result in average detection efficiencies in the range of 50 – 60% with high flatness values of ~1.05. This detection efficiency is not as high as the values demonstrated by the best sate-of-the-art $^3$He tube-base detectors, such as BRIKEN (65 – 70%). The main reason is the energy loss of $^{10}$B(n, α)$^7$Li reaction products in the converter layer.

One quantity which is still to be established is the lower limit of the dark count rate for $^{10}$B-RPCs. Currently there are only data for small-area (<100 cm$^2$) thermal neutron detectors, showing values of 0.02 cps/cm$^2$. Since for such detectors the corresponding dark count rates are negligible, no efforts were made up to date to decrease the background. To be competitive in some applications of fast neutron detectors, such as, for example, the studies of β-delayed neutron emitters, the dark count rate of $^{10}$B-RPCs has to be reduced by an order of magnitude. Several approaches to achieve this have been discussed in this paper.

The combinations of the average detection efficiency and flatness obtained in this study should not be considered the ultimate values. The detector design can be further optimized for each concrete application taking into account the specific performance requirements and the limitations on the total $^{10}$B-RPC area.

Our future plans include design and assembly of a small-scale $^{10}$B-RPC-based fast neutron detector and a comprehensive characterization of the prototype at a neutron source.